\documentclass[fleqn,10pt]{wlscirep}
\usepackage{hyperref}
\usepackage{graphicx,graphics}
\usepackage{amsmath}
\usepackage{wasysym}
\usepackage{csquotes}
\usepackage{cases}
\usepackage{bm}

\title{Efficient Synchronization of Dipolarly Coupled Vortex-Based Spin Transfer Nano-Oscillators}

\author[1,*]{Nicolas Locatelli}
\author[2]{Abbass Hamadeh}
\author[3]{Flavio Abreu Araujo}
\author[4,5]{Anatoly D. Belanovsky}
\author[4,5]{Petr N. Skirdkov}
\author[1]{Romain Lebrun}
\author[1,2,6]{Vladimir V. Naletov}
\author[4,5,7]{Konstantin A. Zvezdin}
\author[8]{Manuel Mu\~{n}oz}
\author[1]{Julie Grollier}
\author[2,+]{Olivier Klein}
\author[1,**]{Vincent Cros}
\author[2]{Grégoire de Loubens}

\affil[1]{Unit\'e Mixte de Physique CNRS/Thales and Universit\'e Paris Sud, F91767 Palaiseau, France}
\affil[2]{Service de Physique de l'Etat Condens\'e, CNRS UMR 3680, CEA Saclay, F91191 Gif-sur-Yvette, France}
\affil[3]{Institute of Condensed Matter and Nanosciences, Universit\'e catholique de Louvain, BE-1348 Louvain-la-Neuve, Belgium}
\affil[4]{Moscow Institute of Physics and Technology (State University), Institutskiy per. 9, 141700 Dolgoprudny, Russia}
\affil[5]{A. M. Prokhorov General Physics Institute, RAS, Vavilova 38, Moscow, Russia}
\affil[6]{Institute of Physics, Kazan Federal University, Kazan 420008, Russian Federation}
\affil[7]{Istituto P.M. srl, Via Grassi 4, Torino, Italy}
\affil[8]{Instituto de Microelectr\'{o}nica de Madrid-IMM (CNM-CSIC), Isaac Newton 8-PTM, 28760 Tres Cantos, Madrid, Spain}

\affil[*]{nicolas.locatelli@u-psud.fr, Present address: Institut d'Electronique Fondamentale, Universit\'e Paris-Sud, UMR CNRS 8622, F91405 Orsay, France}
\affil[+]{Present address: INAC-SPINTEC, CEA/CNRS and Univ. Grenoble Alpes, 38000 Grenoble, France}
\affil[**]{vincent.cros@thalesgroup.com}

\begin{abstract}
Due to their nonlinear properties, spin transfer nano-oscillators can easily adapt their frequency to external stimuli. This makes them interesting model systems to study the effects of synchronization and brings some opportunities to improve their microwave characteristics in view of their applications in information and communication technologies and to design innovative computing architectures. So far, mutual synchronization of spin transfer nano-oscillators through propagating spin-waves and exchange coupling in a common magnetic layer has been demonstrated. Here we show that the dipolar interaction is also an efficient mechanism to synchronize neighbouring oscillators. We experimentally study a pair of vortex-based spin-transfer nano-oscillators, in which mutual synchronization can be achieved despite a significant frequency mismatch between oscillators. Importantly, the coupling efficiency is controlled by the magnetic configuration of the vortices, as confirmed by an analytical model highlighting the physics at play in the synchronization process as well as by micromagnetic simulations.
\end{abstract}

\begin{document}

\flushbottom
\maketitle

\thispagestyle{empty}

\section*{Introduction}

Beyond the traditional applications to data storage and field sensors, the recent progresses in spin transfer physics allows a widening of the application spectra for spintronics devices, notably toward multifunctional devices~\cite{Locatelli_2014_NM_13, Locatelli_2015_DATECED_} relying on their nonvolatile nature, scalability, and compatibility with existing CMOS processes.
Among these novel opportunities, it is anticipated that spin-transfer nano-oscillators (STNO) can represent a breakthrough in the next generation of information and communication technologies. These devices execute self-sustained magnetic oscillations in the GHz range that are induced by the spin-transfer torque and which can be efficiently converted into resistance and voltage oscillations through magneto-resistive effects \cite{Hillebrands_2006_SDCMSI_101}.
To achieve these expectations, STNOs have been intensively studied in the last decade in order to improve the understanding of the mechanisms of the spin transfer torque (STT) as well as the physics of high frequency nonlinear magnetization dynamics~\cite{Hamadeh_2012_PRB_85}.
 
One of the most important properties of STNOs is their nonlinear nature and the associated (natural) tendency to adapt their own frequency to any change in their environment. On one hand, this effect contributes in certain limits to a broadening of the spectral linewidth, deteriorating the quality factor of the STNOs~\cite{Kim_2008_PRL_100}. On the other hand, STNOs appear as model nanoscale systems for studying the effects of synchronization, notably in the regime of large nonlinearities~\cite{Slavin_2009_ITM_45} and even their tendency to reach chaotic regimes under the influence of spin transfer forces \cite{Petit-watelot_2012_NP_8}. Moreover, achieving synchronization of STNOs aims at improving the self-sustained oscillations stability that is crucial for radio-frequency applications as nanoscale tunable radiofrequency source or radiofrequency detector~\cite{Grimaldi_2014_FCSFII_, Prokopenko_2011_IML_2, Stan_2014_IISNAN_}, and enables the development of innovative computing architectures, particularly oscillator-based associative memories based on information coding in the individual phases of large scale arrays of interacting STNOs~\cite{Csaba_2012_TIWCNNTAC_, Levitan_2012_TIWCNNTAC_}.

The forced or non-autonomous dynamics of STNOs has been mainly investigated through the study of their ability to achieve phase locking to an external excitation being an external rf field~\cite{Hamadeh_2012_PRB_85, Urazhdin_2010_PRL_105, Demidov_2014_NC_5, Hamadeh_2014_APL_104} or rf current~ \cite{Rippard_2005_PRL_95, Georges_2008_PRL_101, Quinsat_2011_APL_98, Li_2011_PRB_83, Baraduc_2011_PS_8100, Dussaux_2011_APL_98, Lehndorff_2010_APL_97, Lebrun_2015_PRL_115}. A further advance has been the observation of mutual synchronization between several STNOs. Different coupling mechanisms have been investigated, such as propagating spin-waves in an extended magnetic media~\cite{Kaka_2005_N_437, Mancoff_2005_N_437, Sani_2013_NC_4} or magnetic coupling through exchange interactions~\cite{Ruotolo_2009_NN_4}. However, these approaches to couple STNOs might be arduous for reaching synchronized states in large arrays of STNOs. Two other strategies have been recently investigated theoretically in order to achieve mutual synchronization of a large number of STNOs namely the electric coupling by self-modulation of the current flowing through each device~\cite{Grollier_2006_PRB_73} and the magnetic coupling between the oscillators arising from the dipolar influence of closely packed STNOs~\cite{Belanovsky_2012_PRB_85, Belanovsky_2013_APL_103, Erokhin_2014_PRB_89, Abreuaraujo_2015_PRB_92}. Our main objective is to propose a complete study combining experimental results and numerical as well as theoretical investigation of the mutual synchronization through dipolar coupling of two adjacent vortex-based STNOs. We demonstrate the ability of the two STNOs to achieve mutual synchronization despite a significant frequency mismatch. Moreover, we develop an analytical model describing the dynamics of the synchronization process and highlight the important role of the magnetic parameters (that are the core polarity and vortex chirality of the vortices) of the two STNOs for the optimization of the coupling efficiency. 

\section{Synchronization of two neighbouring vortex-based STNOs}

\subsection*{Experimental results}
\label{sec:ExperimentalEvidence}

\begin{figure}[ht]
\centering
\includegraphics[width=.6\linewidth]{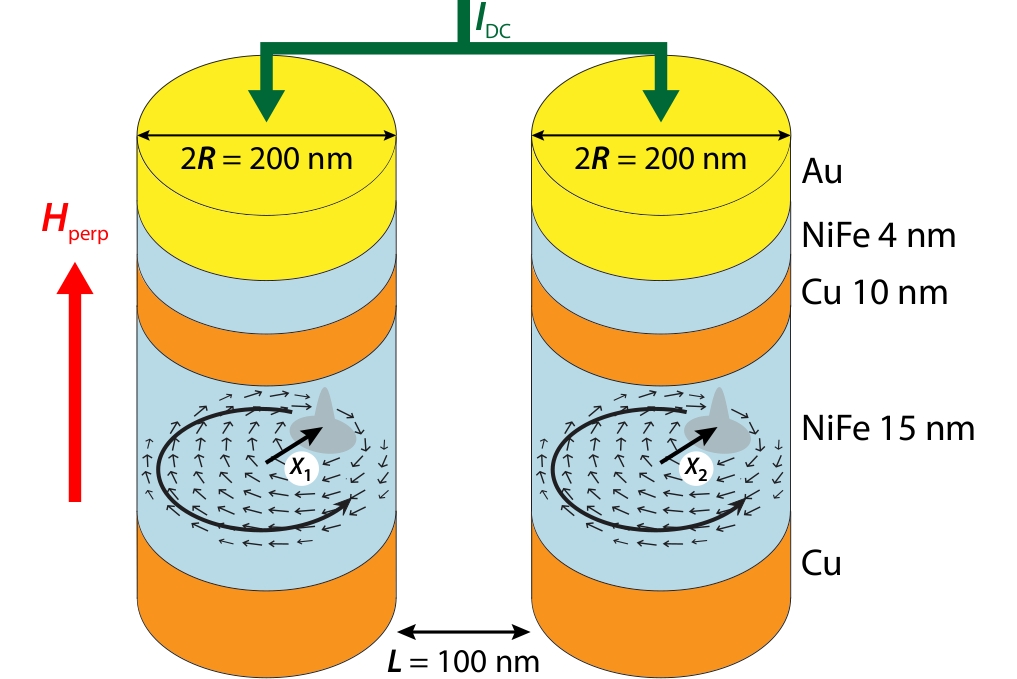}
\caption{Scheme of our coupled STNOs system. Two spin valve nanopillars with $2R=200$nm diameter separated by a distance $L=100$nm. Each pillar contains a NiFe(4nm)/Cu(10nm)/NiFe(15nm) spin valve (NiFe = Ni$_{81}$Fe$_{19}$) and a vortex is nucleated in each 15nm NiFe layers. The current is injected in parallel into the two pillars.}
\label{fig:System}
\end{figure}

Our system is composed of two $2R=200$nm diameter spin valve nanopillars separated by $L=100$nm. These STNOs are patterned by e-beam lithography and ion etching from a magnetic trilayer deposited by sputtering: NiFe(15nm)/Cu(8nm)/NiFe(4nm) (see Fig.~\ref{fig:System}). 
From our previous studies on a single STNO made from the same trilayer stack, we observed that the magnetic configuration in the thicker NiFe layer is a magnetic vortex at remanence. Depending on the applied dc current, the configuration in the thin NiFe layer can be either a quasi-uniform magnetization or a second magnetic vortex~\cite{Locatelli_2011_APL_98, Locatelli_2015_ITM_PP, Abreuaraujo_2012_PRB_86}. In both cases, the resulting spin transfer forces on the thick layer vortex can lead to the sustained excitation of the vortex core gyration if the current sign is appropriately chosen.  In the case of two vortices, an important outcome was to demonstrate that highly coherent gyrotropic oscillations of the coupled vortices can be achieved when their core polarities are in opposite directions~\cite{Locatelli_2011_APL_98, Locatelli_2015_ITM_PP, Abreuaraujo_2012_PRB_86, Abreuaraujo_2013_APL_102}. Further description of the sample can be found in the methods section.

Spintronic oscillators based on the dynamics of a magnetic vortex core have recently proven interest by achieving record quality factors and signal amplitudes~\cite{Pribiag_2007_NP_3, Dussaux_2010_NC_1, Hamadeh_2014_PRL_112}. The main purpose here will be to further consider gyrotropic spin transfer vortex dynamics as a model system to investigate the synchronization of dipolarly coupled oscillators.

The dynamic behaviour of a magnetic vortex core is known to be well described through a single collective variable equation, the so-called Thiele equation describing the vortex core motion through its vector coordinates $\vec{X}$ ~\cite{Thiele_1973_PRL_30, Ivanov_2007_PRL_99, Mistral_2008_PRL_100, Guslienko_2008_JNN_8, Khvalkovskiy_2009_PRB_80, Gaididei_2010_IJQC_110} :

\begin{equation}
\label{ThielEq}
\vec{G}\times\dot{\vec{X}} = - \alpha\eta G\dot{\vec{X}} - k(X) \vec{X} + F_{\text{\text{STT}}}(\vec{X})
\end{equation}

\noindent The gyrotropic force (left hand side of Eq.\ref{ThielEq}) is equal to the sum of the effective dissipative and conservative forces acting on the vortex core, which can be derived from the integration of the local magnetization dynamics. The three forces acting on the vortex core expressed in the right part of Eq.\ref{ThielEq} are: the viscous damping force (proportional and opposed to the core velocity), the spring-like confinement force (opposed to the core displacement), and an additional force accounting for the effect of spin transfer torques. Details of the introduced parameters are given in supplementary informations. To achieve the dynamical regime of self-sustained gyration of the vortex core around the dot centre, the energy provided by the current and the associated spin transfer effect must fully compensate the energy dissipation. Two important conclusions arising from Thiele equation are that (i) the sense of gyration of the vortex core is directly related to the direction of the core polarity~\cite{Guslienko_2008_JNN_8} and (ii) the gyrotropic mode frequency can be tuned by the application of a perpendicular magnetic field $H_{\text{perp}}$ with a quasi-linear evolution: $\omega(H)=\omega_{0}\left(1+P\frac{H_{\text{perp}}}{H_{\text{S}}}\right)$ where $\omega_{0}$ is the zero-field frequency and $H_{\text{S}}$ is the saturation field of the ferromagnet with a slope sign that is directly related to the sign of the vortex polarity $P$~\cite{Deloubens_2009_PRL_102}. In the following, we will use these two properties in order to reliably prepare the system of two neighbouring vortex-based STNOs in two different configurations being either parallel core polarities (Pc) or opposite core polarities (APc) between the two thick layer vortices.

\begin{figure}[ht]
\centering
\includegraphics[width=0.75\linewidth]{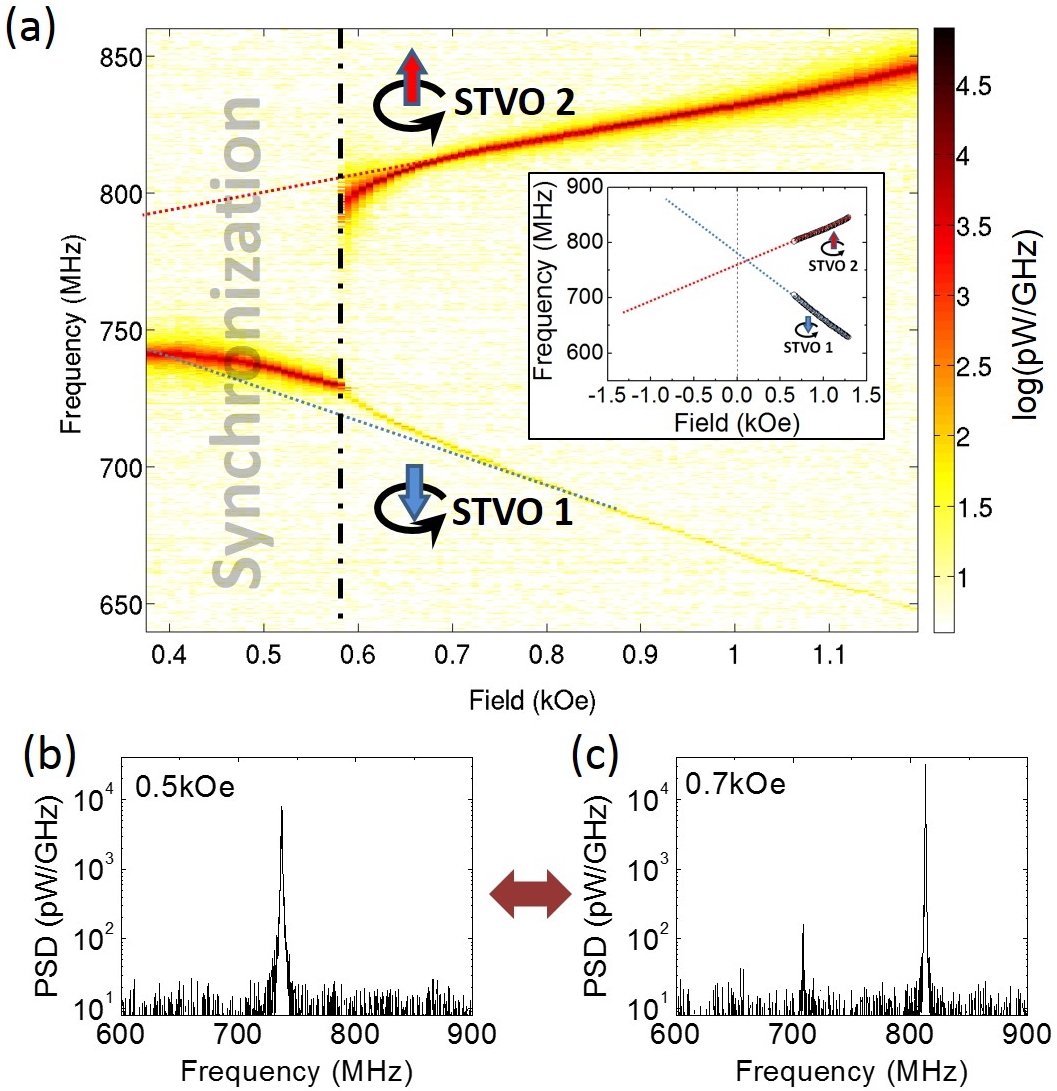}
\caption{(a) Power spectrum map versus perpendicular field measured on Sample \#1, in a case when vortices have opposite core polarities. The injected current through the two pillars is $I_{dc}=50$mA. Insert shows evolution of frequencies versus field for the non-synchronized oscillators and their predicted linear evolution. Power spectra obtained at (b) 0.5 kOe and (c) 0.7 kOe, before (b) and after (c) synchronization is achieved.}
\label{fig:Sample1}
\end{figure}

As shown in Fig.~\ref{fig:System}, the two STNOs are electrically connected in parallel. When the gyrotropic frequencies of the two STNOs are far from each other, we measure two independent peaks on the spectrum analyser, representative of the vortex dynamics of the two STNOs, as illustrated in Fig.~\ref{fig:Sample1}(c). The oscillating in-plane stray field associated with the gyrotropic motion of each vortex will be the source of the dipolar dynamical field used to couple the two STNOs. Contributions of the out-of-plane components of the magnetization to the dynamical coupling are negligible because of the small volume of the vortex cores.

In order to investigate the regime of potential synchronization through dynamical dipolar coupling, we have prepared the system of STNOs in order to have the oscillating vortices with opposite core polarities (APc) which is reflected by the two opposite slopes in their perpendicular field dependence shown in Fig.~\ref{fig:Sample1}(a), measured under a bias current $I_{dc}=+50$mA. 
As shown in Fig.~\ref{fig:Sample1}, when starting from $H_{\text{perp}}$ = 1.2 kOe, the frequency of the two STNOs are about 200MHz apart from each other. This energy difference is very large compared to the coupling energy and the two STNOs almost do not feel each other. Then a mean to tune the frequency mismatch between the two STNOs is to sweep down the perpendicular field. As shown in  Fig.~\ref{fig:Sample1}(c) for $H_{\text{perp}}$ = 0.7kOe, the two STNOs are still behaving almost independently as two separate peaks can be measured. A drastic change is observed for a field slightly smaller field than $H_{\text{perp}}$ = 0.6kOe for which a single peak is recorded (see Fig.~\ref{fig:Sample1}(b)). The observation of such a transition from two peaks associated to each auto-oscillator to a single peak is the evidence of the synchronization of the two STNOs. To discard the possibility that this transition might be related to a magnetic switching (\textit{e.g.}, reversal of a vortex polarity), we have verified that this phenomenon is reversible by sweeping the field back and forth around the critical field value. Indeed, no hysteric transition is observed, as it would be the case if the observed behaviour would be associated with the reversal of the core polarity of one of the two vortices~\cite{Locatelli_2013_APL_102, Abreuaraujo_2012_PRB_86}. Frequency pulling can also be observed, which is characteristic of a transition towards synchronized oscillations~\cite{Pikovsky_2003__}. The measured critical frequency mismatch before synchronization is $\Delta f_{AP}\simeq80$MHz.

Note that the diameters of two STNOs are not perfectly identical (about 3$\%$ difference) which results in two different frequencies at zero applied field as shown in the inset of Fig.~\ref{fig:Sample1} by extrapolating the unsynchronized frequency dependence. This difference in diameters explains why the synchronization is not observed around zero field, since the mode crossing occurs at positive fields $H_{\text{perp}} \simeq$  0.12kOe. It is also noticed that the power amplitudes of the signals associated to each STNO when they are not synchronized are very different. Yet, this difference does not necessarily reflect a strong difference in the vortex gyration amplitudes as the associated voltage oscillations depend strongly on the actual magnetic distribution of the polarizing layers~\cite{Skirdkov_2012_S_02}. 

Moreover, no notable modification of the linewidth was observed, being about $500$kHz before and after the synchronization. This differs from some of the previous observations, where STNOs synchronization was achieved through spin-wave coupling and a clear reduction of the peak linewidth was highlighted~\cite{Kaka_2005_N_437, Mancoff_2005_N_437}.

\begin{figure}[ht]
\centering
	\includegraphics[width=\linewidth]{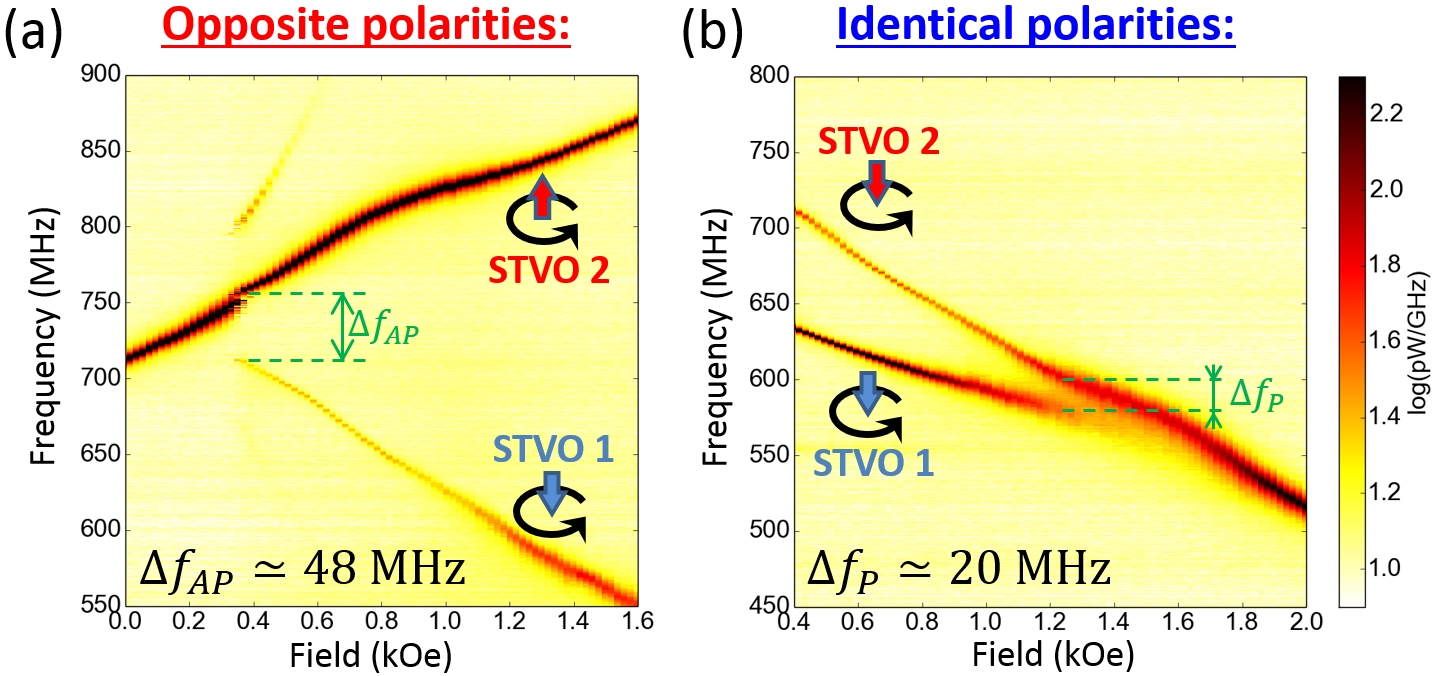}
	\caption{Power spectrum maps versus perpendicular field measured on Sample \#2, in a case when vortices have opposite polarities (a) and when vortices have identical polarities (b). The injected current is $I_{dc}=35$mA.}
\label{fig:Sample2}
\end{figure}

Our method to tune the frequency difference using a perpendicular field is obviously less efficient when the two core polarities are identical (Pc). In order to be able to study the synchronization in case of parallel core polarity, we have measured another pair of STNOs which have more separate natural gyrotropic frequencies. As shown in Fig.~\ref{fig:Sample2}(b), a reversible transition from independent to synchronized auto-oscillations is observed. Note that the maximum frequency mismatch before synchronization is about $\Delta f_{P}=20\pm2$MHz for these measurements with identical core polarities. In order to make a quantitative comparison, we have also measured on the same systems of coupled STNOs the synchronization in case of opposite core polarities (APc).  As shown in figure~\ref{fig:Sample2}(a), for the same bias current $I_{dc}=+35$mA, we find that the maximum frequency mismatch is about $\Delta f_{AP}=48\pm2$MHz. On this measurement, the appearance of a secondary peak around $800$MHz before synchronization is again characteristic of the interaction of the two oscillators, corresponding to the frequency beating phenomenon~\cite{Pikovsky_2003__}.

A striking result is thus that $\Delta f_{AP}>\Delta f_{P}$, suggesting that the coupling energy between STNOs is more efficient to induce synchronization when the two interacting vortices have opposite core polarities. It is important to emphasize that the two cases mostly differ through the relative direction of gyration of the two vortices: the two vortex cores have the same sense of gyration when $P_{1}P_{2}>0$, whereas they gyrate in opposite directions when $P_{1}P_{2}<0$.

It is interesting to note that after synchronization, the resulting frequency and associated dependence to $H_{\text{perp}}$ of the synchronized oscillations are close to one of the two auto-oscillators. Indeed, in each presented measurement, we can identify a \textit{leader} oscillator, and a \textit{follower}. We interpret that the latter has a stronger ability to adapt its frequency to external stimuli than the leader, and then experiences a bigger frequency shift to achieve synchronization.

\subsection*{Dipolar interaction between the oscillators}
\label{sec:DipolarCoupling}

To establish a model, we consider a system composed of the two thick layers' vortices, interacting through the mutual dipolar fields created by their rotating in-plane net magnetizations~\cite{Shibata_2003_PRB_67, Jung_2010_APL_97, Sukhostavets_2011_APE_4, Sugimoto_2011_PRL_106, Belanovsky_2012_PRB_85, Belanovsky_2013_APL_103, Abreuaraujo_2015_PRB_92}. Contributions of the out-of-plane components of the magnetization to the coupling are neglected because of the small volume of the vortex cores. Given the large positive current flowing through each pillar, the two vortices have identical chiralities following the direction of the current-induced Oersted field. Moreover, given the smaller volume of the thin $4$nm layers, the possible influence of shallow dynamics occurring in these layers will not be considered and these polarizing layers' magnetizations will be considered fixed. 

We consider that the net in-plane magnetization of the shifted vortex increases linearly with the core displacement from the dot centre: $\left<\vec{M}_{i}\right>/M_{\text{S}}=C_{i}\xi \hat{z}\times \vec{X}_{i}$, where $\xi\simeq 2/3$ is determined analytically~\cite{Guslienko_2002_JAP_91}, and $\hat{z}$ is oriented along the pillar axis. While the vortex is gyrating, its net in-plane magnetization describes a $360^{\circ}$ rotation. To model the dipolar coupling between the two vortices, we consider a simple model of two rotating in-plane macro-dipoles $\vec{M}_{1}$ and $\vec{M}_{2}$ at the disks' centre positions~\cite{Belanovsky_2012_PRB_85, Belanovsky_2013_APL_103, Abreuaraujo_2015_PRB_92}. The interaction energy can then be expressed as:

\begin{eqnarray}
	W_{\text{int}} &=& \frac{\mu_{0}}{4\pi D_{12}^{3}} \left( \vec{M}_{1}\cdot\vec{M}_{2} - 3(\vec{M_{1}}\cdot\vec{e}_{12})(\vec{M_{2}}\cdot\vec{e}_{12}) \right) \\
		&=&  C_{1}C_{2} \left(\mu_{(+)}X_{1} X_{2} \cos(\varphi_{1}+\varphi_{2}) - \mu_{(-)} X_{1} X_{2} \cos(\varphi_{1}-\varphi_{2}) \right)
	\label{eq:WintDipoles}
\end{eqnarray}

\noindent where $\vec{X}_{1}(X_{1},\varphi_{1})$ and $\vec{X}_{2}(X_{2},\varphi_{2})$ are the polar coordinates of the $1^{st}$ and $2^{nd}$ vortices cores respectively, and $\vec{e}_{12}$ is a unit vector in the direction joining the two dots centres (Fig.~\ref{fig:System}). The coupling coefficients can then be rewritten as $\mu_{(-)}=\frac{1}{2}\frac{\xi^{2}\mu_{0}{M_{\text{S}}}^{2}V^{2}}{4\pi D_{12}^{3}}$ and $\mu_{(+)}=\frac{3}{2}\frac{\xi^{2}\mu_{0}{M_{\text{S}}}^{2}V^{2}}{4\pi D_{12}^{3}}$ so that $\mu_{(-)}=3\mu_{(+)}$, with $D_{12}=2R+L$ the interpillar centre-to-centre distance. In the following, the influence of the perpendicular field $H_{\text{perp}}$ on these two coupling parameters will be neglected, considering that the applied field is a small portion of the saturating field ($H_{\text{S}}\simeq9$kOe), so that the vortices are not significantly tilted out-of-plane.

The vortex core direction of gyration is directly related to its polarity, which is translated in polar coordinates by: $\dot{\varphi_{i}}=P_{i}\omega_{i}$ where $\omega_{i}$ is the absolute gyrotropic frequency of the vortex. As a consequence, the term relating to $\mu_{(+)}$ should oscillate at the frequency $(P_{1}\omega_{1}+P_{2}\omega_{2})$, while the second term relating to $\mu_{(-)}$ should oscillate at the frequency $(P_{1}\omega_{1}-P_{2}\omega_{2})$. The variation rates of the two terms depend then strongly on the relative polarities of the gyrating vortices. 

\subsection*{Analytical model for synchronization}
\label{sec:AnalyticalModel}

The purpose is now to define the synchronization criterion. We propose here to model the coupled system through two coupled Thiele equations:

\begin{subnumcases}{}
	\vec{G}_{1}\times\dot{\vec{X}}_{1} = - \alpha\eta_{1}G_{1} \dot{\vec{X}}_{1} - k_{1}(X_{1})\vec{X}_{1} + \vec{F}_{\text{STT}}(\vec{X}_{1}) + \vec{F}_{\text{int}}^{2 \rightarrow 1}(\vec{X}_{2}) \\
	\vec{G}_{2}\times\dot{\vec{X}}_{2} = - \alpha\eta_{2}G_{2} \dot{\vec{X}}_{2} - k_{2}(X_{2})\vec{X}_{2} + \vec{F}_{\text{STT}}(\vec{X}_{2}) + \vec{F}_{\text{int}}^{1 \rightarrow 2}(\vec{X}_{1})
\end{subnumcases}

\noindent where the interaction force is then expressed as : $\vec{F}_{\text{int}}^{j \rightarrow i}=-\frac{\partial W_{\text{int}}}{\partial \vec{X}_{i}}$. In the general case of a non-uniform polarizer, with magnetization distributed in plane or out-of-plane, the spin transfer force can be described by the general expression: $\vec{F}_{\text{STT}}(\vec{X}_{i})=P_{i}\lambda(X_{i},J_{i})\hat{z}\times \vec{X}_{i}$ where $J_{i}$ is the current density flowing through the pillar and $\lambda(X_{i},J_{i})$ expresses the spin transfer efficiency~\cite{Gaididei_2010_IJQC_110, Ivanov_2007_PRL_99, Khvalkovskiy_2009_PRB_80, Khvalkovskiy_2010_APL_96, Mistral_2008_PRL_100, Sluka_2012_PRB_86}. By using complex number coordinates, these equations can then be written as:

\begin{subnumcases}{}
	\begin{aligned}
	\frac{\dot{X_{i}}}{X_{i}} = \left[ \frac{\lambda(X_{i},J_{i})}{G_{i}} - \frac{\alpha\eta_{i}k_{i}(X_{i})}{G_{i}} \right] & - \frac{\alpha\eta_{i} C_{1}C_{2}}{G_{i}} \text{Re}\left[ \mu_{(+)} \frac{\bm{X}_{j}^{*}}{\bm{X}_{i}} - \mu_{(-)} \frac{\bm{X}_{j}}{\bm{X}_{i}} \right] \\
	& - P_{i} \frac{C_{1}C_{2}}{G_{i}}  \text{Im}\left[ \mu_{(+)} \frac{\bm{X}_{j}^{*}}{\bm{X}_{i}} - \mu_{(-)} \frac{\bm{X}_{j}}{\bm{X}_{i}} \right]  
	\end{aligned} \\
	\begin{aligned}
	\dot{\varphi_{i}} = P_{i} \left[ \frac{k_{i}(X_{i})}{G_{i}} + \frac{\alpha\eta_{i}\lambda(X_{i},J_{i})}{G_{i}} \right] & - \frac{\alpha\eta_{i} C_{1}C_{2}}{G_{i}} \text{Im}\left[ \mu_{(+)} \frac{\bm{X}_{j}^{*}}{\bm{X}_{i}} - \mu_{(-)} \frac{\bm{X}_{j}}{\bm{X}_{i}} \right] \\
	& + P_{i} \frac{C_{1}C_{2}}{G_{i}} \text{Re}\left[ \mu_{(+)} \frac{\bm{X}_{j}^{*}}{\bm{X}_{i}} - \mu_{(-)} \frac{\bm{X}_{j}}{\bm{X}_{i}} \right]
	\end{aligned}
\end{subnumcases}

\noindent where $\bm{X}_{i}=X_{i}e^{\mathrm{i}\varphi_{i}}$ ($i\in\left\{1,2\right\}$) are the complex coordinates of the core positions and ${\bm{X}_{1,2}}^{*}=X_{i}e^{-\mathrm{i}\varphi_{i}}$ are their complex conjugates.

The coupling action is considered as a perturbation to the equilibrium auto-oscillations of the isolated pillars. In order to extract a differential equation for the relative phase dynamics, we expand these equations around their equilibrium values through the oscillations power: $p_{i}=\left(\frac{X_{i}}{R_{i}}\right)^{2}=p_{i_{0}}+\delta p_{i}$, and the subsequent deviations of the instantaneous frequencies: $P_{i}\dot{\varphi_{i}} = \omega_{i} + N_{i} \delta p_{i}$, and dissipative forces: $\frac{\alpha\eta_{i}k_{i}(X_{i})}{G_{i}} - \frac{\lambda(X_{i},J_{i})}{G_{i}} = 2 \Gamma_{p_{i}} \delta p_{i}$ where $N_{i}$'s are the coefficients of the non-linear frequency shifts, and $\Gamma_{p_{i}}$'s are the damping rates for small power deviations \cite{Slavin_2009_ITM_45}.

Details of the developments of these equations are proposed in supplementary informations. To capture the dominant mechanisms responsible for synchronization, the following approximations are used: $G_{1} \simeq G_{2}$, $\eta_{1} \simeq \eta_{2}$, $p_{1_{0}} \simeq p_{2_{0}} $, $N_{1} \simeq N_{2}$, $\Gamma_{p_{1}} \simeq \Gamma_{p_{2}}$, given the small applied field and the small differences between the two pillars. We further suppose that the two auto-oscillators only differ through their frequencies $\omega_{1} \neq \omega_{2}$. We then extract the differential equation governing the dynamics of the signed phase difference $\Psi=(P_{1}\varphi_{1}-P_{2}\varphi_{2})$, in the cases of identical polarities ($P_{1}=P_{2}=+1$) and opposite polarities ($P_{1} = -P_{2}$):
\begin{itemize}
	\item if $P_{1}=P_{2}$: $\Psi=\varphi_{1}-\varphi_{2}$
	\begin{equation}
		\dot{\Psi} = (\omega_{1} - \omega_{2}) - 2 C_{1}C_{2} \frac{\mu_{(-)}}{G} \left( \nu + \alpha\eta \right) \sin(\Psi) \label{eq:Adler-like1}
	\end{equation}
	
	\item if $P_{1}=-P_{2}$: $\Psi=\varphi_{1}+\varphi_{2}$
	\begin{equation}
		\dot{\Psi} = (\omega_{1} - \omega_{2}) + 2 C_{1}C_{2} \frac{\mu_{(+)}}{G} \left( \nu + \alpha\eta \right) \sin(\Psi) \label{eq:Adler-like2}
	\end{equation}
\end{itemize}
\noindent where $\nu_{i}=\frac{N_{i}p_{i_{0}}}{\Gamma_{p_{i}}}$ is the normalized dimensionless nonlinear frequency shift.
Equations~\ref{eq:Adler-like1} and \ref{eq:Adler-like2} are similar to the typical Adler equation~ \cite{Pikovsky_2003__}. Synchronization can be achieved only if a stable solution exists for this equation. Given that $C_{1}=C_{2}$ in our experiment, the conditions on the frequency mismatch $\Delta\omega$ for synchronization are:
\begin{subnumcases}{\label{eq:Sync}}
	\Delta\omega_{P} < \frac{2 \mu_{(-)}}{G} \left( \nu + \alpha\eta \right) & for $P_{1}=P_{2}$ \\
	\Delta\omega_{AP} < \frac{2 \mu_{(+)}}{G} \left( \nu + \alpha\eta \right) & for $P_{1}=-P_{2}$
\end{subnumcases}

Equation~\ref{eq:Sync} states that there are two mechanisms contributing to synchronization between the two STNOs. The first mechanism, associated with the $\alpha\eta$ multiplying coefficient, corresponds to the direct action of the dipolar force on the oscillator phase. The second mechanism corresponds to the action of the dipolar force leading to the modification of the vortex orbits and subsequently their gyration frequency to achieve synchronization. 
For the STNOs under study, we evaluate that $\alpha\eta\simeq0.01$ and $\nu\apprge 1$ \cite{Hamadeh_2014_PRL_112}, highlighting that the second mechanism is clearly the dominant one.

From this analytical study, it appears that the coupling action efficiency depends on the relative polarities. Notably, from the expression of dipolar coupling presented in Eq.~\ref{eq:WintDipoles}, this model predicts that when polarities are anti-parallel (APc) the synchronization can occur for a frequency mismatch that is three times larger than when polarities are parallel (Pc): $\Delta\omega_{AP}/\Delta\omega_{P} = \mu_{(+)}/\mu_{(-)}=3$. 

These results are in good agreement with the experimental observations we presented. Indeed, the critical frequency mismatches before synchronization measured on the second sample verify $\Delta f_{AP}/\Delta f_{P}=48
MHz/20MHz\simeq2.4$. This confirms that the effective coupling energy is stronger when the vortices are gyrating in opposite directions, and that this configuration is preferable to achieve efficient synchronization.

Finally, another prediction from the model is that, depending on the relative polarities, the equilibrium phase shift after synchronization is different, being either close to $\Psi=\varphi_{1}-\varphi_{2}=0$ for identical core polarities or to $\Psi=\varphi_{1}+\varphi_{2}=\pi$ for opposite core polarities. Changing the relative chiralities signs would reverse this statement, but without affecting the coupling amplitude and the associated synchronization features.

Note that in Eq.~\ref{eq:Sync}, we neglected interaction terms associated to fast variations of the coupling energy which do not participate in the synchronization process. However, it should be noted that the high frequency oscillations of the coupling energy still exist and are certainly perturbing the synchronized oscillations. Notably, a major consequence will be to increase the phase fluctuations of the synchronized oscillators, thus increasing the synchronized signal linewidth. This phenomenon might prevent the synchronization phenomenon to increase the coherence of the oscillations.

\begin{figure}[ht]
\centering
	\includegraphics[width=.7\linewidth]{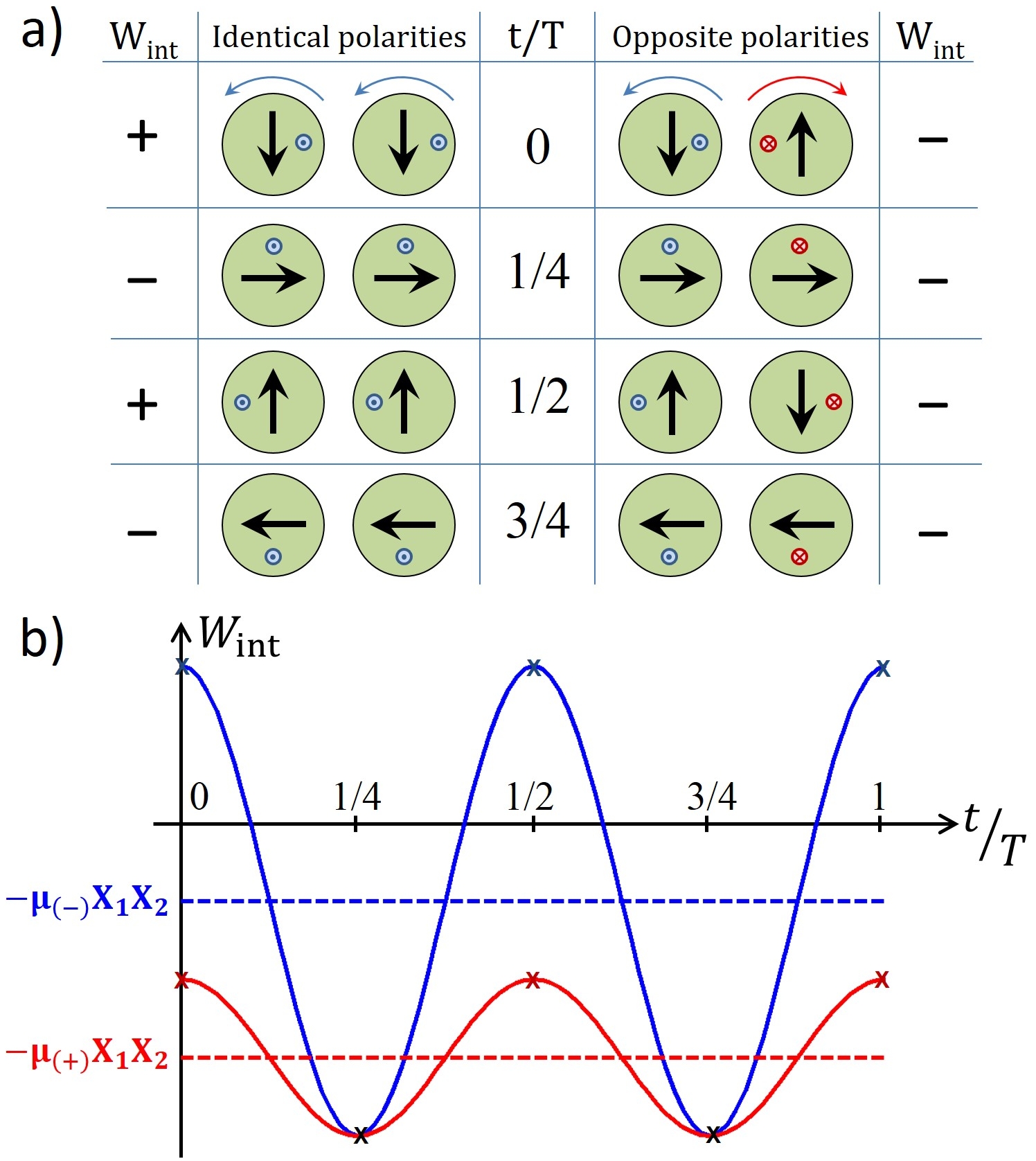}
	\caption{(a) Illustration of the synchronized motion of the two auto-oscillating vortices cores, and of their net in-plane magnetizations (black arrows), for the cases of identical polarities (left) and opposite polarities (right) during one period. The sign of the associated dipolar coupling energy $W_{\text{int}}$ is also given. (b) Predicted evolution of the dipolar coupling energy $W_{\text{int}}$ versus part of oscillation period for identical (blue) and opposite polarities (red). The dotted lines show the average interaction energy in both cases.}
\label{fig:Dipolar}
\end{figure}

\subsection*{Study of the dipolar energy versus relative polarities}

To gain further insight on the relative efficiency of the dipolar action for synchronization depending on the relative polarities, we present in figure~\ref{fig:Dipolar} a sketch of the temporal evolution of the coupling energy in the ideal condition of phase locking $\omega_{1}=\omega_{2}$ for the two cases of identical polarities ($\varphi_{2}=\varphi_{1}$) and opposite polarities ($\varphi_{2}=\pi-\varphi_{1}$). In the case of opposite polarities, when vortices gyrate in opposite directions, the relative alignment of the net magnetizations is always favourable during the motion, so that the mean interaction energy is large and negative. In contrast, when vortices gyrate in identical direction (polarities are identical), the synchronized gyration brings the system in situations when net magnetizations are in a favourable alignment ($W_{\text{int}}<0$) but also in a unfavourable alignment ($W_{\text{int}}>0$).

To summarize, in case of identical polarities, $W_{\text{int}}$ oscillates with large amplitude with a mean energy $\left.\left<W_{\text{int}}\right>\right|_{P}=-\mu_{(-)}X_{1}X_{2}$, while in case of opposite polarities, $W_{\text{int}}$ oscillates with small amplitude but with $\left.\left<W_{\text{int}}\right>\right|_{AP}=-\mu_{(+)}X_{1}X_{2}$, three times larger. In other words, the dipolar interaction is more efficient for stabilizing synchronization when in AP configuration.

\begin{figure}[ht]
\centering
	\includegraphics[width=.7\linewidth]{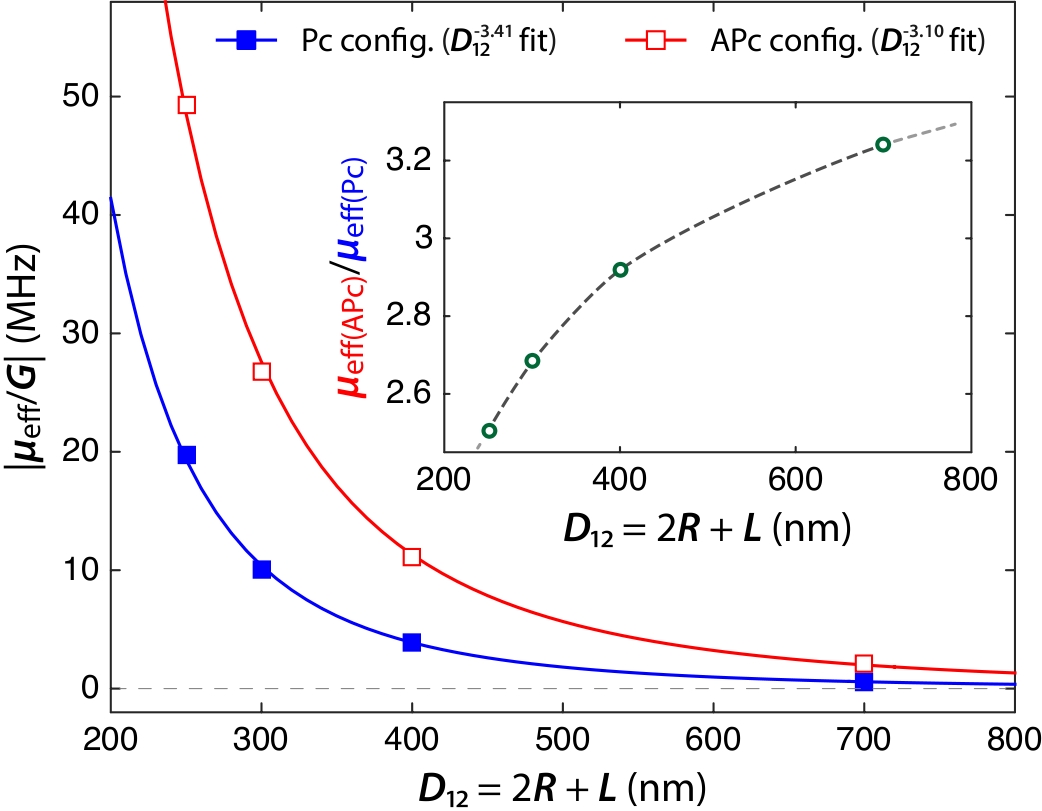}
	\caption{Average interaction energy versus interpillar distance extracted from micromagnetic simulations for the case of two identical synchronized oscillators with radii $R=100$nm, when vortices have identical polarities (blue line and filled squares) and opposite polarities (red line and open squares).}
\label{fig:DipolarEnergy}
\end{figure}

Finally, the validity of the macro-dipoles and Thiele approach was evaluated by conducting a micromagnetic study. Using full micromagnetic simulations, we studied the synchronization dynamics between the two identical oscillators, with no frequency mismatch, at zero temperature and zero field. This specific case allowed us to monitor the synchronization phenomenon whatever the amplitude of the coupling. Following the approach described in \cite{Belanovsky_2012_PRB_85, Belanovsky_2013_APL_103, Abreuaraujo_2015_PRB_92}, we extracted the effective coupling coefficient $\mu_{\text{eff}}$, defined as $\left.\left<W_{\text{int}}\right>\right|=-\mu_{\text{eff}}X_{1}X_{2}$, as a function of the centre-to-centre distance. The result is presented in Fig.\ref{fig:DipolarEnergy}, where the ratio $\mu_{\text{eff}}/G$ is plotted for the two polarities configurations. This study allows us to get a quantitative prediction of the critical frequency mismatch for synchronization as a function of the pillars interdistance.

It appears from this study that the prediction $\mu_{\text{eff}}(\text{APc})/\mu_{\text{eff}}(\text{Pc})\sim 3$ is a fair approximation in the present experimental case where $D_{12}=300$nm. Yet, this ratio (see insert in Fig.\ref{fig:DipolarEnergy}) is predicted to decrease as the pillars interdistance decreases. Comparison to experimental results suggests that the real interpillar distance is actually smaller than the nominal one.

\section*{Discussion}

In this study, both experiments and analytical modelling demonstrated that the effective dipolar coupling energy, and the subsequent critical frequency mismatch for synchronization, is stronger for STNOs with interacting vortices gyrating in opposite direction i.e., with opposite polarities, than for interacting vortices gyrating in the same direction i.e., with identical polarities. Synchronization of STNOs with frequency mismatches corresponding, respectively, to approximately $10\%$ and $6\%$ of the mean frequency of the two oscillators have been observed in this optimal configuration. These numbers fall reasonably in the range of observed frequency spreading in networks of STNOs, and the results can be extrapolated for potential synchronization of a large number of oscillators.
Applications of such networks have been proposed recently for the realization of neuro-inspired STNO-based associative memories \cite{Csaba_2012_TIWCNNTAC_, Levitan_2012_TIWCNNTAC_}. In such architectures, the control over the coupling between neighbouring oscillators is a major asset as it will be the support of the information. Considering the implementation of the coupling between oscillators, these results can be safely extended to other spin-transfer oscillators architectures, such as the ones based on magnetic tunnel junctions, which recently demonstrated higher power with low linewidth signal~\cite{Dussaux_2010_NC_1, Dussaux_2014_APL_105}.

\section*{Methods}

\label{supp:Samples}

The samples are composed of two $2R=200$nm circular spin-valve nanopillars separated by a $L=100$nm edge-to-edge distance, patterned in a Cu(60nm)/NiFe(15nm)/Cu(10nm)/NiFe(4nm)/Au(25nm) stack (NiFe=Ni$_{81}$Fe$_{19}$) by standard e-beam and ion etching lithography process. 

Samples with single pillars prepared from the same stack have already been intensively studied in previous works~\cite{Locatelli_2011_APL_98,Locatelli_2013_APL_102,Hamadeh_2014_PRL_112,Hamadeh_2014_APL_104,Locatelli_2015_ITM_PP}. Applying a positive current $I_{dc}>10$mA ensures that the two vortices have identical positive chilarities (direction of the curling magnetization) following the direction of the current induced Oersted field. Control over the vortices polarities can be done by applying a perpendicular magnetic field~\cite{Locatelli_2013_APL_102}.

We note that the induced Oersted field from each pillar is also felt by the neighbouring pillar. This crosstalk field was evaluated between $100$Oe and $200$Oe for a current $I_{dc}=20$mA flowing in each pillar. The consequence will be an offset of the vortex cores equilibrium positions by a few nanometers, being much smaller than the pillar radius. However, this field is not sufficient to affect the vortex stability.

The two self-sustained oscillators are supplied in parallel through single top and bottom electrodes for current injection (see figure~\ref{fig:System}), and the amplitudes of the current flowing through each pillar are hence not independently controlled. Despite possible small size and shape deviations between the two pillars, we assume that equal current flows through each pillar. In our convention, a positive current corresponds to electrons flowing from the thick to the thin layer. Given the parallel supply, it is not possible to independently measure the magnetoresistive signals associated to each pillar, but only to measure the voltage across the two pillars.

The ac-voltage is isolated from biasing circuit by a bias-T and then measured on a spectrum analyzer after a 30dB amplification.


\begin{thebibliography}{10}
\expandafter\ifx\csname url\endcsname\relax
  \def\url#1{\texttt{#1}}\fi
\expandafter\ifx\csname urlprefix\endcsname\relax\def\urlprefix{URL }\fi
\providecommand{\bibinfo}[2]{#2}
\providecommand{\eprint}[2][]{\url{#2}}

\bibitem{Locatelli_2014_NM_13}
\bibinfo{author}{Locatelli, N.}, \bibinfo{author}{Cros, V.} \&
  \bibinfo{author}{Grollier, J.}
\newblock \bibinfo{title}{Spin-torque building blocks}.
\newblock \emph{\bibinfo{journal}{Nat. Mater.}} \textbf{\bibinfo{volume}{13}},
  \bibinfo{pages}{11--20} (\bibinfo{year}{2014}).

\bibitem{Locatelli_2015_DATECED_}
\bibinfo{author}{Locatelli, N.} \emph{et~al.}
\newblock \bibinfo{title}{Spintronic devices as key elements for
  energy-efficient neuroinspired architectures}.
\newblock In \emph{\bibinfo{booktitle}{Design, Automation Test in Europe
  Conference Exhibition {(DATE)}, 2015}}, \bibinfo{pages}{994--999}
  (\bibinfo{year}{2015}).

\bibitem{Hillebrands_2006_SDCMSI_101}
\bibinfo{author}{Stiles, M.~D.} \& \bibinfo{author}{Miltat, J.}
\newblock \bibinfo{title}{{Spin-Transfer} torque and dynamics}.
\newblock In \bibinfo{editor}{Hillebrands, B.} \& \bibinfo{editor}{Thiaville,
  A.} (eds.) \emph{\bibinfo{booktitle}{Spin Dynamics in Confined Magnetic
  Structures {III}}}, vol. \bibinfo{volume}{101}, \bibinfo{pages}{225--308}
  (\bibinfo{publisher}{Springer Berlin Heidelberg}, \bibinfo{year}{2006}).

\bibitem{Hamadeh_2012_PRB_85}
\bibinfo{author}{Hamadeh, A.} \emph{et~al.}
\newblock \bibinfo{title}{Autonomous and forced dynamics in a spin-transfer
  nano-oscillator: Quantitative magnetic-resonance force microscopy}.
\newblock \emph{\bibinfo{journal}{Phys. Rev. B}} \textbf{\bibinfo{volume}{85}},
  \bibinfo{pages}{140408} (\bibinfo{year}{2012}).

\bibitem{Kim_2008_PRL_100}
\bibinfo{author}{Kim, J.}, \bibinfo{author}{Tiberkevich, V.} \&
  \bibinfo{author}{Slavin, A.~N.}
\newblock \bibinfo{title}{Generation linewidth of an {Auto-Oscillator} with a
  nonlinear frequency shift: {Spin-Torque} {Nano-Oscillator}}.
\newblock \emph{\bibinfo{journal}{Phys. Rev. Lett.}}
  \textbf{\bibinfo{volume}{100}}, \bibinfo{pages}{017207}
  (\bibinfo{year}{2008}).

\bibitem{Slavin_2009_ITM_45}
\bibinfo{author}{Slavin, A.} \& \bibinfo{author}{Tiberkevich, V.}
\newblock \bibinfo{title}{Nonlinear {Auto-Oscillator} theory of microwave
  generation by {Spin-Polarized} current}.
\newblock \emph{\bibinfo{journal}{IEEE Trans. Magn.}}
  \textbf{\bibinfo{volume}{45}}, \bibinfo{pages}{1875--1918}
  (\bibinfo{year}{2009}).

\bibitem{Petit-watelot_2012_NP_8}
\bibinfo{author}{{Petit-Watelot}, S.} \emph{et~al.}
\newblock \bibinfo{title}{Commensurability and chaos in magnetic vortex
  oscillations}.
\newblock \emph{\bibinfo{journal}{Nat. Phys.}} \textbf{\bibinfo{volume}{8}},
  \bibinfo{pages}{682--687} (\bibinfo{year}{2012}).

\bibitem{Grimaldi_2014_FCSFII_}
\bibinfo{author}{Grimaldi, E.} \emph{et~al.}
\newblock \bibinfo{title}{Spintronic nano-oscillators: Towards nanoscale and
  tunable frequency devices}.
\newblock In \emph{\bibinfo{booktitle}{Frequency Control Symposium {(FCS)},
  2014 {IEEE} International}}, \bibinfo{pages}{1--6} (\bibinfo{year}{2014}).

\bibitem{Prokopenko_2011_IML_2}
\bibinfo{author}{Prokopenko, O.}, \bibinfo{author}{Bankowski, E.},
  \bibinfo{author}{Meitzler, T.}, \bibinfo{author}{Tiberkevich, V.} \&
  \bibinfo{author}{Slavin, A.}
\newblock \bibinfo{title}{{Spin-Torque} {Nano-Oscillator} as a microwave signal
  source}.
\newblock \emph{\bibinfo{journal}{IEEE Magn. Lett.}}
  \textbf{\bibinfo{volume}{2}}, \bibinfo{pages}{3000104--3000104}
  (\bibinfo{year}{2011}).

\bibitem{Stan_2014_IISNAN_}
\bibinfo{author}{Stan, M.}, \bibinfo{author}{Kabir, M.}, \bibinfo{author}{Wolf,
  S.} \& \bibinfo{author}{Lu, J.}
\newblock \bibinfo{title}{Spin torque nano oscillators as key building blocks
  for the {Systems-on-Chip} of the future}.
\newblock In \emph{\bibinfo{booktitle}{2014 {IEEE/ACM} International Symposium
  on Nanoscale Architectures {(NANOARCH)}}}, \bibinfo{pages}{37--38}
  (\bibinfo{year}{2014}).

\bibitem{Csaba_2012_TIWCNNTAC_}
\bibinfo{author}{Csaba, G.} \emph{et~al.}
\newblock \bibinfo{title}{Spin torque oscillator models for applications in
  associative memories}.
\newblock In \emph{\bibinfo{booktitle}{2012 13th International Workshop on
  Cellular Nanoscale Networks and Their Applications {(CNNA)}}},
  \bibinfo{pages}{1--2} (\bibinfo{year}{2012}).

\bibitem{Levitan_2012_TIWCNNTAC_}
\bibinfo{author}{Levitan, S.} \emph{et~al.}
\newblock \bibinfo{title}{{Non-Boolean} associative architectures based on
  nano-oscillators}.
\newblock In \emph{\bibinfo{booktitle}{2012 13th International Workshop on
  Cellular Nanoscale Networks and Their Applications {(CNNA)}}},
  \bibinfo{pages}{1--6} (\bibinfo{year}{2012}).

\bibitem{Urazhdin_2010_PRL_105}
\bibinfo{author}{Urazhdin, S.}, \bibinfo{author}{Tabor, P.},
  \bibinfo{author}{Tiberkevich, V.} \& \bibinfo{author}{Slavin, A.}
\newblock \bibinfo{title}{Fractional synchronization of {Spin-Torque}
  {Nano-Oscillators}}.
\newblock \emph{\bibinfo{journal}{Phys. Rev. Lett.}}
  \textbf{\bibinfo{volume}{105}}, \bibinfo{pages}{104101}
  (\bibinfo{year}{2010}).

\bibitem{Demidov_2014_NC_5}
\bibinfo{author}{Demidov, V.~E.} \emph{et~al.}
\newblock \bibinfo{title}{Synchronization of spin hall nano-oscillators to
  external microwave signals}.
\newblock \emph{\bibinfo{journal}{Nat. Commun.}} \textbf{\bibinfo{volume}{5}},
  \bibinfo{pages}{3179} (\bibinfo{year}{2014}).

\bibitem{Hamadeh_2014_APL_104}
\bibinfo{author}{Hamadeh, A.} \emph{et~al.}
\newblock \bibinfo{title}{Perfect and robust phase-locking of a spin transfer
  vortex nano-oscillator to an external microwave source}.
\newblock \emph{\bibinfo{journal}{Appl. Phys. Lett.}}
  \textbf{\bibinfo{volume}{104}}, \bibinfo{pages}{022408}
  (\bibinfo{year}{2014}).

\bibitem{Rippard_2005_PRL_95}
\bibinfo{author}{Rippard, W.~H.} \emph{et~al.}
\newblock \bibinfo{title}{Injection locking and phase control of spin transfer
  nano-oscillators}.
\newblock \emph{\bibinfo{journal}{Phys. Rev. Lett.}}
  \textbf{\bibinfo{volume}{95}}, \bibinfo{pages}{067203}
  (\bibinfo{year}{2005}).

\bibitem{Georges_2008_PRL_101}
\bibinfo{author}{Georges, B.} \emph{et~al.}
\newblock \bibinfo{title}{Coupling efficiency for phase locking of a spin
  transfer {Nano-Oscillator} to a microwave current}.
\newblock \emph{\bibinfo{journal}{Phys. Rev. Lett.}}
  \textbf{\bibinfo{volume}{101}}, \bibinfo{pages}{017201}
  (\bibinfo{year}{2008}).

\bibitem{Quinsat_2011_APL_98}
\bibinfo{author}{Quinsat, M.} \emph{et~al.}
\newblock \bibinfo{title}{Injection locking of tunnel junction oscillators to a
  microwave current}.
\newblock \emph{\bibinfo{journal}{Appl. Phys. Lett.}}
  \textbf{\bibinfo{volume}{98}}, \bibinfo{pages}{182503--182503--3}
  (\bibinfo{year}{2011}).

\bibitem{Li_2011_PRB_83}
\bibinfo{author}{Li, D.}, \bibinfo{author}{Zhou, Y.}, \bibinfo{author}{Zhou,
  C.} \& \bibinfo{author}{Hu, B.}
\newblock \bibinfo{title}{Fractional locking of spin-torque oscillator by
  injected ac current}.
\newblock \emph{\bibinfo{journal}{Phys. Rev. B}} \textbf{\bibinfo{volume}{83}},
  \bibinfo{pages}{174424} (\bibinfo{year}{2011}).

\bibitem{Baraduc_2011_PS_8100}
\bibinfo{author}{Baraduc, C.} \emph{et~al.}
\newblock \bibinfo{title}{Synchronization of high power vortex oscillators at
  multiple of the fundamental frequency}.
\newblock \emph{\bibinfo{journal}{Proc. SPIE}} \textbf{\bibinfo{volume}{8100}},
  \bibinfo{pages}{810016--810016--6} (\bibinfo{year}{2011}).

\bibitem{Dussaux_2011_APL_98}
\bibinfo{author}{Dussaux, A.} \emph{et~al.}
\newblock \bibinfo{title}{Phase locking of vortex based spin transfer
  oscillators to a microwave current}.
\newblock \emph{\bibinfo{journal}{Appl. Phys. Lett.}}
  \textbf{\bibinfo{volume}{98}}, \bibinfo{pages}{132506--132506--3}
  (\bibinfo{year}{2011}).

\bibitem{Lehndorff_2010_APL_97}
\bibinfo{author}{Lehndorff, R.}, \bibinfo{author}{B\"{u}rgler, D.~E.},
  \bibinfo{author}{Schneider, C.~M.} \& \bibinfo{author}{Celinski, Z.}
\newblock \bibinfo{title}{Injection locking of the gyrotropic vortex motion in
  a nanopillar}.
\newblock \emph{\bibinfo{journal}{Appl. Phys. Lett.}}
  \textbf{\bibinfo{volume}{97}}, \bibinfo{pages}{142503--142503--3}
  (\bibinfo{year}{2010}).

\bibitem{Lebrun_2015_PRL_115}
\bibinfo{author}{Lebrun, R.} \emph{et~al.}
\newblock \bibinfo{title}{Understanding of phase noise squeezing under
  fractional synchronization of a nonlinear spin transfer vortex oscillator}.
\newblock \emph{\bibinfo{journal}{Phys. Rev. Lett.}}
  \textbf{\bibinfo{volume}{115}}, \bibinfo{pages}{017201}
  (\bibinfo{year}{2015}).

\bibitem{Kaka_2005_N_437}
\bibinfo{author}{Kaka, S.} \emph{et~al.}
\newblock \bibinfo{title}{Mutual phase-locking of microwave spin torque
  nano-oscillators}.
\newblock \emph{\bibinfo{journal}{Nature}} \textbf{\bibinfo{volume}{437}},
  \bibinfo{pages}{389--392} (\bibinfo{year}{2005}).

\bibitem{Mancoff_2005_N_437}
\bibinfo{author}{Mancoff, F.~B.}, \bibinfo{author}{Rizzo, N.~D.},
  \bibinfo{author}{Engel, B.~N.} \& \bibinfo{author}{Tehrani, S.}
\newblock \bibinfo{title}{Phase-locking in double-point-contact spin-transfer
  devices}.
\newblock \emph{\bibinfo{journal}{Nature}} \textbf{\bibinfo{volume}{437}},
  \bibinfo{pages}{393--395} (\bibinfo{year}{2005}).

\bibitem{Sani_2013_NC_4}
\bibinfo{author}{Sani, S.} \emph{et~al.}
\newblock \bibinfo{title}{Mutually synchronized bottom-up multi-nanocontact
  spin{\textendash}torque oscillators}.
\newblock \emph{\bibinfo{journal}{Nat. Commun.}} \textbf{\bibinfo{volume}{4}}
  (\bibinfo{year}{2013}).

\bibitem{Ruotolo_2009_NN_4}
\bibinfo{author}{Ruotolo, A.} \emph{et~al.}
\newblock \bibinfo{title}{Phase-locking of magnetic vortices mediated by
  antivortices}.
\newblock \emph{\bibinfo{journal}{Nat. Nanotechnol.}}
  \textbf{\bibinfo{volume}{4}}, \bibinfo{pages}{528--532}
  (\bibinfo{year}{2009}).

\bibitem{Grollier_2006_PRB_73}
\bibinfo{author}{Grollier, J.}, \bibinfo{author}{Cros, V.} \&
  \bibinfo{author}{Fert, A.}
\newblock \bibinfo{title}{Synchronization of spin-transfer oscillators driven
  by stimulated microwave currents}.
\newblock \emph{\bibinfo{journal}{Phys. Rev. B}} \textbf{\bibinfo{volume}{73}},
  \bibinfo{pages}{060409} (\bibinfo{year}{2006}).

\bibitem{Belanovsky_2012_PRB_85}
\bibinfo{author}{Belanovsky, A.~D.} \emph{et~al.}
\newblock \bibinfo{title}{Phase locking dynamics of dipolarly coupled
  vortex-based spin transfer oscillators}.
\newblock \emph{\bibinfo{journal}{Phys. Rev. B}} \textbf{\bibinfo{volume}{85}},
  \bibinfo{pages}{100409} (\bibinfo{year}{2012}).

\bibitem{Belanovsky_2013_APL_103}
\bibinfo{author}{Belanovsky, A.~D.} \emph{et~al.}
\newblock \bibinfo{title}{Numerical and analytical investigation of the
  synchronization of dipolarly coupled vortex spin-torque nano-oscillators}.
\newblock \emph{\bibinfo{journal}{Appl. Phys. Lett.}}
  \textbf{\bibinfo{volume}{103}}, \bibinfo{pages}{122405}
  (\bibinfo{year}{2013}).

\bibitem{Erokhin_2014_PRB_89}
\bibinfo{author}{Erokhin, S.} \& \bibinfo{author}{Berkov, D.}
\newblock \bibinfo{title}{Robust synchronization of an arbitrary number of
  spin-torque-driven vortex nano-oscillators}.
\newblock \emph{\bibinfo{journal}{Phys. Rev. B}} \textbf{\bibinfo{volume}{89}},
  \bibinfo{pages}{144421} (\bibinfo{year}{2014}).

\bibitem{Abreuaraujo_2015_PRB_92}
\bibinfo{author}{Abreu~Araujo, F.} \emph{et~al.}
\newblock \bibinfo{title}{Optimizing magnetodipolar interactions for
  synchronizing vortex based spin-torque nano-oscillators}.
\newblock \emph{\bibinfo{journal}{Phys. Rev. B}} \textbf{\bibinfo{volume}{92}},
  \bibinfo{pages}{045419} (\bibinfo{year}{2015}).

\bibitem{Locatelli_2011_APL_98}
\bibinfo{author}{Locatelli, N.} \emph{et~al.}
\newblock \bibinfo{title}{Dynamics of two coupled vortices in a spin valve
  nanopillar excited by spin transfer torque}.
\newblock \emph{\bibinfo{journal}{Appl. Phys. Lett.}}
  \textbf{\bibinfo{volume}{98}}, \bibinfo{pages}{062501--062501--3}
  (\bibinfo{year}{2011}).

\bibitem{Locatelli_2015_ITM_PP}
\bibinfo{author}{Locatelli, N.} \emph{et~al.}
\newblock \bibinfo{title}{Improved spectral stability in spin transfer
  nano-oscillators: single vortex versus coupled vortices dynamics.}
\newblock \emph{\bibinfo{journal}{IEEE Trans. Magn.}}
  \textbf{\bibinfo{volume}{{PP}}}, \bibinfo{pages}{1--1}
  (\bibinfo{year}{2015}).

\bibitem{Abreuaraujo_2012_PRB_86}
\bibinfo{author}{Abreu~Araujo, F.} \emph{et~al.}
\newblock \bibinfo{title}{Microwave signal emission in spin-torque vortex
  oscillators in metallic nanowires: Experimental measurements and
  micromagnetic numerical study}.
\newblock \emph{\bibinfo{journal}{Phys. Rev. B}} \textbf{\bibinfo{volume}{86}},
  \bibinfo{pages}{064424} (\bibinfo{year}{2012}).

\bibitem{Abreuaraujo_2013_APL_102}
\bibinfo{author}{Abreu~Araujo, F.}, \bibinfo{author}{Piraux, L.},
  \bibinfo{author}{Antohe, V.~A.}, \bibinfo{author}{Cros, V.} \&
  \bibinfo{author}{Gence, L.}
\newblock \bibinfo{title}{Single spin-torque vortex oscillator using combined
  bottom-up approach and e-beam lithography}.
\newblock \emph{\bibinfo{journal}{Appl. Phys. Lett.}}
  \textbf{\bibinfo{volume}{102}}, \bibinfo{pages}{222402}
  (\bibinfo{year}{2013}).

\bibitem{Pribiag_2007_NP_3}
\bibinfo{author}{Pribiag, V.~S.} \emph{et~al.}
\newblock \bibinfo{title}{Magnetic vortex oscillator driven by d.c.
  spin-polarized current}.
\newblock \emph{\bibinfo{journal}{Nat. Phys.}} \textbf{\bibinfo{volume}{3}},
  \bibinfo{pages}{498--503} (\bibinfo{year}{2007}).

\bibitem{Dussaux_2010_NC_1}
\bibinfo{author}{Dussaux, A.} \emph{et~al.}
\newblock \bibinfo{title}{Large microwave generation from current-driven
  magnetic vortex oscillators in magnetic tunnel junctions}.
\newblock \emph{\bibinfo{journal}{Nat. Commun.}} \textbf{\bibinfo{volume}{1}},
  \bibinfo{pages}{8} (\bibinfo{year}{2010}).

\bibitem{Hamadeh_2014_PRL_112}
\bibinfo{author}{Hamadeh, A.} \emph{et~al.}
\newblock \bibinfo{title}{Origin of spectral purity and tuning sensitivity in a
  spin transfer vortex {Nano-Oscillator}}.
\newblock \emph{\bibinfo{journal}{Phys. Rev. Lett.}}
  \textbf{\bibinfo{volume}{112}}, \bibinfo{pages}{257201}
  (\bibinfo{year}{2014}).

\bibitem{Thiele_1973_PRL_30}
\bibinfo{author}{Thiele, A.~A.}
\newblock \bibinfo{title}{{Steady-State} motion of magnetic domains}.
\newblock \emph{\bibinfo{journal}{Phys. Rev. Lett.}}
  \textbf{\bibinfo{volume}{30}}, \bibinfo{pages}{230--233}
  (\bibinfo{year}{1973}).

\bibitem{Ivanov_2007_PRL_99}
\bibinfo{author}{Ivanov, B.~A.} \& \bibinfo{author}{Zaspel, C.~E.}
\newblock \bibinfo{title}{Excitation of spin dynamics by {Spin-Polarized}
  current in vortex state magnetic disks}.
\newblock \emph{\bibinfo{journal}{Phys. Rev. Lett.}}
  \textbf{\bibinfo{volume}{99}}, \bibinfo{pages}{247208}
  (\bibinfo{year}{2007}).

\bibitem{Mistral_2008_PRL_100}
\bibinfo{author}{Mistral, Q.} \emph{et~al.}
\newblock \bibinfo{title}{{Current-Driven} vortex oscillations in metallic
  nanocontacts}.
\newblock \emph{\bibinfo{journal}{Phys. Rev. Lett.}}
  \textbf{\bibinfo{volume}{100}}, \bibinfo{pages}{257201}
  (\bibinfo{year}{2008}).

\bibitem{Guslienko_2008_JNN_8}
\bibinfo{author}{Guslienko, K.~Y.}
\newblock \bibinfo{title}{Magnetic vortex state stability, reversal and
  dynamics in restricted geometries}.
\newblock \emph{\bibinfo{journal}{J Nanosci. Nanotechno.}}
  \textbf{\bibinfo{volume}{8}}, \bibinfo{pages}{2745--2760}
  (\bibinfo{year}{2008}).

\bibitem{Khvalkovskiy_2009_PRB_80}
\bibinfo{author}{Khvalkovskiy, A.~V.}, \bibinfo{author}{Grollier, J.},
  \bibinfo{author}{Dussaux, A.}, \bibinfo{author}{Zvezdin, K.~A.} \&
  \bibinfo{author}{Cros, V.}
\newblock \bibinfo{title}{Vortex oscillations induced by spin-polarized current
  in a magnetic nanopillar: Analytical versus micromagnetic calculations}.
\newblock \emph{\bibinfo{journal}{Phys. Rev. B}} \textbf{\bibinfo{volume}{80}},
  \bibinfo{pages}{140401} (\bibinfo{year}{2009}).

\bibitem{Gaididei_2010_IJQC_110}
\bibinfo{author}{Gaididei, Y.}, \bibinfo{author}{Kravchuk, V.~P.} \&
  \bibinfo{author}{Sheka, D.~D.}
\newblock \bibinfo{title}{Magnetic vortex dynamics induced by an electrical
  current}.
\newblock \emph{\bibinfo{journal}{Int. J. Quant. Chem.}}
  \textbf{\bibinfo{volume}{110}}, \bibinfo{pages}{83--97}
  (\bibinfo{year}{2010}).

\bibitem{Deloubens_2009_PRL_102}
\bibinfo{author}{de~Loubens, G.} \emph{et~al.}
\newblock \bibinfo{title}{Bistability of vortex core dynamics in a single
  perpendicularly magnetized nanodisk}.
\newblock \emph{\bibinfo{journal}{Phys. Rev. Lett.}}
  \textbf{\bibinfo{volume}{102}}, \bibinfo{pages}{177602}
  (\bibinfo{year}{2009}).

\bibitem{Locatelli_2013_APL_102}
\bibinfo{author}{Locatelli, N.} \emph{et~al.}
\newblock \bibinfo{title}{Reversal process of a magnetic vortex core under the
  combined action of a perpendicular field and spin transfer torque}.
\newblock \emph{\bibinfo{journal}{Appl. Phys. Lett.}}
  \textbf{\bibinfo{volume}{102}}, \bibinfo{pages}{062401}
  (\bibinfo{year}{2013}).

\bibitem{Pikovsky_2003__}
\bibinfo{author}{Pikovsky, A.}, \bibinfo{author}{Kurths, J.} \&
  \bibinfo{author}{Rosenblum, M.}
\newblock \emph{\bibinfo{title}{Synchronization : a universal concept in
  nonlinear sciences}} (\bibinfo{publisher}{Cambridge Univ. Press},
  \bibinfo{address}{Cambridge [u.a.]}, \bibinfo{year}{2003}).

\bibitem{Skirdkov_2012_S_02}
\bibinfo{author}{Skirdkov, P.~N.} \emph{et~al.}
\newblock \bibinfo{title}{Influence of shape imperfection on dynamics of vortex
  {Spin-Torque} {Nano-Oscillator}}.
\newblock \emph{\bibinfo{journal}{{SPIN}}} \textbf{\bibinfo{volume}{02}},
  \bibinfo{pages}{1250005--1} (\bibinfo{year}{2012}).

\bibitem{Shibata_2003_PRB_67}
\bibinfo{author}{Shibata, J.}, \bibinfo{author}{Shigeto, K.} \&
  \bibinfo{author}{Otani, Y.}
\newblock \bibinfo{title}{Dynamics of magnetostatically coupled vortices in
  magnetic nanodisks}.
\newblock \emph{\bibinfo{journal}{Phys. Rev. B}} \textbf{\bibinfo{volume}{67}},
  \bibinfo{pages}{224404} (\bibinfo{year}{2003}).

\bibitem{Jung_2010_APL_97}
\bibinfo{author}{Jung, H.} \emph{et~al.}
\newblock \bibinfo{title}{Observation of coupled vortex gyrations by
  70-ps-time- and 20-nm-space-resolved full-field magnetic transmission soft
  x-ray microscopy}.
\newblock \emph{\bibinfo{journal}{Appl. Phys. Lett.}}
  \textbf{\bibinfo{volume}{97}}, \bibinfo{pages}{222502--222502--3}
  (\bibinfo{year}{2010}).

\bibitem{Sukhostavets_2011_APE_4}
\bibinfo{author}{Sukhostavets, O.~V.}, \bibinfo{author}{Gonzalez, J.~M.} \&
  \bibinfo{author}{Guslienko, K.~Y.}
\newblock \bibinfo{title}{Magnetic vortex excitation frequencies and eigenmodes
  in a pair of coupled circular dots}.
\newblock \emph{\bibinfo{journal}{Appl. Phys. Express}}
  \textbf{\bibinfo{volume}{4}}, \bibinfo{pages}{065003} (\bibinfo{year}{2011}).

\bibitem{Sugimoto_2011_PRL_106}
\bibinfo{author}{Sugimoto, S.} \emph{et~al.}
\newblock \bibinfo{title}{Dynamics of coupled vortices in a pair of
  ferromagnetic disks}.
\newblock \emph{\bibinfo{journal}{Phys. Rev. Lett.}}
  \textbf{\bibinfo{volume}{106}}, \bibinfo{pages}{197203}
  (\bibinfo{year}{2011}).

\bibitem{Guslienko_2002_JAP_91}
\bibinfo{author}{Guslienko, K.~Y.} \emph{et~al.}
\newblock \bibinfo{title}{Eigenfrequencies of vortex state excitations in
  magnetic submicron-size disks}.
\newblock \emph{\bibinfo{journal}{J. Appl. Phys.}}
  \textbf{\bibinfo{volume}{91}}, \bibinfo{pages}{8037--8039}
  (\bibinfo{year}{2002}).

\bibitem{Khvalkovskiy_2010_APL_96}
\bibinfo{author}{Khvalkovskiy, A.~V.} \emph{et~al.}
\newblock \bibinfo{title}{Nonuniformity of a planar polarizer for
  spin-transfer-induced vortex oscillations at zero field}.
\newblock \emph{\bibinfo{journal}{Appl. Phys. Lett.}}
  \textbf{\bibinfo{volume}{96}}, \bibinfo{pages}{212507--212507--3}
  (\bibinfo{year}{2010}).

\bibitem{Sluka_2012_PRB_86}
\bibinfo{author}{Sluka, V.} \emph{et~al.}
\newblock \bibinfo{title}{Quenched slonczewski windmill in spin-torque vortex
  oscillators}.
\newblock \emph{\bibinfo{journal}{Phys. Rev. B}} \textbf{\bibinfo{volume}{86}},
  \bibinfo{pages}{214422} (\bibinfo{year}{2012}).

\bibitem{Dussaux_2014_APL_105}
\bibinfo{author}{Dussaux, A.} \emph{et~al.}
\newblock \bibinfo{title}{Large amplitude spin torque vortex oscillations at
  zero external field using a perpendicular spin polarizer}.
\newblock \emph{\bibinfo{journal}{Appl. Phys. Lett.}}
  \textbf{\bibinfo{volume}{105}}, \bibinfo{pages}{022404}
  (\bibinfo{year}{2014}).

\end{thebibliography}

\section*{Acknowledgements}
The authors acknowledge the ANR agency (SPINNOVA ANR-11-NANO-0016) as well as EU FP7 grant (MOSAIC No. ICT-FP7-8.317950) for financial support. RFBR grant 10-02-01162 is acknowledged. A.D.B. acknowleges the Dynasty Foundation for the financial support. V.V.N. acknowledges support from program Competitive Growth of KFU. F.A.A. acknowledges the Research Science Foundation of Belgium (FRS-FNRS) for financial support (FRIA grant) and the UCL for an FSR complement (Fonds Spécial de Recherche). A.D.B., P.N.S. and K.A.Z. acknowledge 50 Labs Initiative of Moscow Institute of Physics and Technology. The authors acknowledge Prof. Anatoly Zvezdin for fruitful discussion.

\subsection*{Author contributions statement}

NL, GdL, JG, OK, VC conceived and coordinated the project. NL, AH, VVN and GdL performed the experimental measurements with the help of JG, OK and VC for the interpretation of the data. JG, MM designed and fabricated the samples. FAA, ADB, KAZ performed and analyzed the micromagnetic simulations. NL, FAA, ADB, PNS, RL and KAZ worked on the development of the analytical model. The manuscript was prepared by NL with the assistance of GdL and VC. All authors commented the manuscript.

\end{document}